\documentstyle[11pt,menu97,epsfig]{article}

\begin{document}
\setlength{\baselineskip}{2.6ex}

\title{ Precision Low--Energy Measurements at DA$\Phi$NE\\
and Their Impact on $\bar{K}N$ Physics\thanks{Research supported by the 
E.E.C. Human Capital and Mobility Program under contract No. 
CHRX--CT92--0026.}}

\author{Paolo M. Gensini, Rafael Hurtado\thanks{On leave of absence from 
Centro Internacional de F\'\i{s}ica, Bogot\'a, Colombia, under a grant by 
Colciencias.}\\
{\em Dip. Fisica, Univ. Perugia, and Sez. I.N.F.N., Perugia, Italy}\\
and\\ Galileo Violini\\
{\em Dip. Fisica, Univ. Calabria, and Gr. Coll. I.N.F.N., Cosenza, Italy\\
and\\
Univ. de El Salvador, San Salvador}}

\maketitle

\begin{abstract}
\setlength{\baselineskip}{2.6ex}
We delineate a ``theoretical'' $\chi^2$--functional technique 
to implement fixed--$t$ analyticity on the invariant amplitudes for the 
$\bar{K}N$, $\pi\Lambda$, $\pi\Sigma$ coupled channels in a fast and 
straightforward way.
\end{abstract}

\setlength{\baselineskip}{2.6ex}

\section*{1. Introduction.}

A dedicated scattering experiment at the Frascati $\phi$--factory DA$\Phi$NE 
is expected to bring the statistics on each of the $\bar{K}N$--initiated 
processes from the few hundred events of the TST Collaboration at NIMROD to 
the 10$^5$ -- 10$^6$ events/year expected for a {\em dedicated} $\bar{K}N$ 
experiment at DA$\Phi$NE. To analyse these data with adequate accuracy, 
one has to go beyond the coupled--channel analyses performed in the past. 

With DA$\Phi$NE under construction, and still now that the date of its 
commissioning is nearing, a team of theorists grouped in an E.E.C. Network 
under the name of EuroDA$\Phi$NE to study both widely and deeply the impact 
and the exploitation of the data soon to come out of Frascati: the outputs of 
their work are to be summarized in successive editions of the ``DA$\Phi$NE 
Physics Handbook'' \cite{DHb95}, now to its second version. Of this group the 
present authors are proud to be a small branch.

\section*{2. Physics at DA$\Phi$NE.}

When DA$\Phi$NE will be working at the $\phi$--resonance peak, it will be an 
intense source of almost monochromatic kaons, $K^{\pm}$ of momentum 126 MeV/c 
and $K_L$ of momentum 110 MeV/c (at the interaction point). Depending on the 
nature of the beam--pipe window and on the use of suitably thin moderators, 
the $K^{\pm}$ momenta can be made to span a momentum range from around 115 
MeV/c down to about 80 MeV/c, covering the $\bar{K^0}n$ charge--exchange 
threshold region.

Rates at the source are expected (for a reference luminosity of 
$5\cdot10^{32}$ cm$^{-2}$s$^{-1}$) to be around 1,200 $K^{\pm}$/s and 830 
$K_L$/s: the decay length of the $K^{\pm}$'s being about 90 cm, the volume 
(and cost) of a {\em dedicated} detector, measuring interactions in gaseous 
H$_2$ (and D$_2$ and $^3$He/$^4$He as well), and exploiting KLOE's detection 
techniques, need not be larger that 1/10 of KLOE, but, of course, it will 
have to be as transparent to low--momentum particles and as efficient in 
detecting neutrals as KLOE. Measurements on $^4$He will allow access to the 
kinematical region below the $\bar{K}N$ threshold, dominated by the two 
resonant states $\Lambda$(1405) and $\Sigma$(1385) \cite{Gen95}.

One can thus hope to measure not only differential cross sections for all 
charged and neutral two--body final states, but also channel cross sections 
for all three-body ones and, even more important for the knowledge 
of the low--energy P--waves (whose relevance to the S = -- 1 channels 
phenomenology is treated in the other paper presented by our group at this 
Symposium \cite{Gen97}), 
one can exploit the self--analysing nature of $\Lambda$ and $\Sigma^+$ 
hyperons to measure their final polarizations \cite{Gen95}.

As of today, the only available pieces of information on the P--waves below 
250 MeV/c are 4 data points on the 1st moment of the angular distribution 
for each of the channels $\pi^{\pm}\Sigma^{\mp}$, published by the TST 
Collaboration \cite{Cib82} (and not used by any analysis), all consistent with 
zero within 2 $\sigma$.

\section*{3. Outline of an ``advanced'' coupled--channel analysis.}

An analysis of the new data will of course have to re--analyse all 
existing, lower--quality data as well: for a significant analysis below 
550 MeV/c (the region covered by J.K. Kim \cite{Kim67}) one has 
to use at least four partial waves for each isospin channel; this means a 
total of 36 two--channel K-- or M--matrix elements, requiring thus (in a 
self--consistent amplitude expansion to order $p^4$) more than a hundred 
parametres, since i) one must describe the D$_{03}$ resonance close to 
1,520 MeV c.m. energy, and ii) the inelastic channels are not negligible 
for an accurate analysis.

Non--experimental information is therefore needed to reduce the ambiguities 
of this multi--channel analysis, with information available only on part of 
a single row/column of the coupled--channel T--matrix. In his S--wave 
analysis \cite{Mar81}, A.D.~Martin used once--subtracted forward dispersion 
relations for the amplitudes $C(\omega) = A(\omega) + \omega B(\omega)$ 
(in terms of the 
invariant--amplitude decomposition of the pseudoscalar--baryon T--matrix $T = 
A(\omega) + B(\omega) \cdot \gamma_{\mu}Q^{\mu}$) for both $K^{\pm}p$ and 
$K^{\pm}n$ elastic 
scattering; his constraints depended of course on additional parameters, 
namely the subtraction constants $C(0)$ and the $KNY$ couplings 
$G_{KN\Lambda}^2$ and $G_{KN\Sigma}^2$. One would 
like to extend this type of constraints to other amplitudes and channels, 
as e.g. the $B$ amplitudes \cite{Gen97}, 
dominated by P--waves rather than by S--waves as $C$'s are, or those 
``unphysical'' channels such as $\pi{Y}\to\pi{Y'}$ or $\bar{K}N\to\pi{Y}$, 
with $Y, Y' = \Lambda, \Sigma$.

A first difficulty is encountered in this kind of approach, since 
the K--matrix parametrization is unitary, but only approximately analytic 
on the r.h. cut, since it does not in general possess the l.h. cut 
singularities: this affects the description of r.h. cut 
unphysical regions for reactions $\bar{K}N\to\pi\Sigma$ and 
$\pi\Sigma\to\pi\Sigma$, when the $u$--channel Born--term singularities 
``invade'' the low--energy $\pi\Lambda$ cut. The trouble is only marginal, 
since in the {\em full amplitudes} the singularities reduce to poles present 
only in the real parts, and the imaginary parts generated by K--matrices {\em 
automatically} satisfy unitarity: the answer to this problem, rather that to 
modify the K--matrices, is to make use {\em only} of the imaginary 
parts produced by this formalism in the dispersive treatment of 
these channels.

Conventional forms of dispersion relations \cite{Que74,Isa94}, even in their 
simplest, unsubtracted version \cite{Gen97}, present several shortcomings, 
all of them tied either to their rate of convergence, which could be 
improved by subtractions (but at the cost of additional, {\em external} 
parametres), or to the Born--terms, which introduce the coupling constants, 
which one would like to {\em extract} from the analysis, as further {\em 
external} parametres of the latter.

A way to perform a parametre--free test of analyticity {\em alone} on such 
analyses would be to make direct use of Cauchy's theorem for the amplitude 
$F(z)$, analytic inside a contour $\Gamma$: 
\begin{equation}
\frac{1}{2\pi i} \oint_{\Gamma} F(z) dz = 0 .
\label{1}
\end{equation}

Thus, since our $F$'s have the Born--term poles, we have to write instead 
\begin{equation}
\frac{1}{2\pi i} \oint_{\Gamma} F(z) G(z) dz = 0 ,
\label{2}
\end{equation}
where $G(z)$ is analytic by construction inside $\Gamma$, and 
has zeros at all poles of $F(z)$. Unfortunately, as long as one keeps $z = 
\omega^2$, there is no {\em elementary} way of doing it and still be able to 
work without divergent integrations. 

A way out was indicated by several authors \cite{Ciu70,Nen70}, and it consists 
in mapping the variable $\omega$ into a new one $z(\omega^2)$ so that the 
physical sheet is mapped into the unit disk and the first (or second) 
unphysical one into the rest of the plane: one can now have a family of $G$'s 
with a very simple form, such as $G_n(z) = (z - z_{\infty})^2 \prod_i 
(z - z_i) \cdot z^n$, valid for any crossing--even amplitude $F(\omega^2)$ 
behaving as $\beta \cdot \omega^{\alpha}$ with $1 \ge \alpha > 0$ for $\omega 
\to \infty$, where $z_{\infty} = \lim_{\omega\to\infty} z(\omega^2)$ and the 
$z_i$ represent the positions of the poles inside the unit disk. The mapping 
$z(\omega^2)$ is given as 
\begin{equation}
z(\omega^2) = \frac{\sqrt{\omega_t^2-\omega_0^2}-\sqrt{\omega_t^2-\omega^2}}
{\sqrt{\omega_t^2-\omega_0^2}+\sqrt{\omega_t^2-\omega^2}} ,
\label{3}
\end{equation}
where $\omega_t$ is the threshold of the unphysical sheet mapped outside the 
unit disk and $\omega_0$ is the point in the $\omega$--plane mapped into the 
centre of the same.

Defining 
\begin{equation}
I_n(F) = \frac{1}{2\pi i} \oint_{\Gamma} F(z) G_n(z) dz ,
\label{4}
\end{equation}
one would have all $I_n=0$ for an $F$ meromorphic inside $\Gamma$, and 
building the sum
\begin{equation}
\chi^2_a[F,1] = \sum_{n=0}^N\vert\frac{1}{2\pi i} \int_{\Gamma_1} F(z) 
G_n(z) dz \vert^2 ,
\label{5}
\end{equation}
where $\Gamma_1$ is the part of $\Gamma$ over which the analysis is 
performed (note that the map of Eq. (3) reduces $\Gamma_2 = \Gamma - 
\Gamma_1$ to a small arclet around $z_{\infty} = - 1$); by this $\chi^2$ 
functional one can introduce in the fitting procedure a measure of the 
{\em lack of analyticity} of the parametrization employed. Furthermore, with 
$\omega_t$ chosen for $\pi\Sigma\to\pi\Sigma$ at the $\pi\Sigma$ threshold, 
only the imaginary parts of the amplitudes on the $\pi\Lambda$ low--energy 
cut will enter the integrations.

\section*{4. Further refinements and applications of the method.}

Further improvements to the above technique can be offered by the weighing 
technique, which consists in introducing as a factor in the dispersive 
integrals a function $w(z)$, analytic inside $\Gamma$, which further 
suppresses the effects of the neglected region $\Gamma_2$, while at the 
same time weighing the inputs according to their statistical (or systematic) 
accuracies \cite{Ciu70,Nen70,Ciu75}.

Let $F$ be the correct amplitude, $\hat{F}$ our parametrization 
for it or (when $F$ is directly measurable) its experimental values, and 
let us assume $\vert\hat{F}-F\vert\cdot\vert{G_n}\vert = \vert\hat{F}-F 
\vert\cdot\vert G_0\vert < \epsilon$ on $\Gamma_1$ and $\vert{F}\vert\cdot 
\vert{G_n}\vert = \vert{F}\vert\cdot\vert{G_0}\vert < M$ on $\Gamma_2$: using 
the Schwarz--Villat formula \cite{Ciu70,Nen70,Ciu75,Gen77} 
\begin{equation}
w(z) = \exp \frac{1}{2\pi i} \oint_{\Gamma} \log \vert w(\zeta) \vert 
\frac{\zeta + z}{\zeta - z} \frac{d\zeta}{\zeta}
\label{6}
\end{equation}
one can construct a weight function $w(z)$ satisfying $\vert w(z) \vert = 
\lambda/\epsilon$ on $\Gamma_1$ and $\vert w(z) \vert = \lambda/M$ on 
$\Gamma_2$, and thus define a ``weighted'' theoretical $\chi^2$--functional as 
\begin{equation}
\chi^2_a[F,w] = \sum_{n=0}^N\vert\frac{1}{2\pi i} \int_{\Gamma_1} F(z) 
G_n(z) w(z) dz \vert^2.
\label{7}
\end{equation}

Such a test can then be run 
\begin{itemize}
\begin{description}
\item{i)} for all invariant amplitudes $A$, $B$ and $C$;
\item{ii)} in a reasonably wide range of $t$ values, such as 
$\vert t \vert \le 2 m_{\pi}^2$;
\item{iii)} for five out of the ten processes between low--mass, S = - 1 
two--body isospin eigenchannels (of which only nine involved in the analysis).
\end{description}
\end{itemize}

The above method can indeed cover both $\bar{K}N\to\bar{K}N$ I = 0, 1 isospin 
channels (using also the input from the $KN$ ones), 
the $2F_1 - F_0$ crossing--even 
combination of $\pi\Sigma\to\pi\Sigma$ amplitudes (the lower index 
standing for the $s$--channel isospin), the $\pi\Lambda\to\pi\Lambda$ and the 
$\pi\Lambda\to\pi\Sigma$ ones (which are pure isospin 1).

With the same method one could easily handle these amplitudes to extract 
from them their low--energy parametres in the unphysical region \cite{Gen77} 
around $\omega^2 = 0$ in the range $\vert t \vert \le 2 m_{\pi}^2$: this 
includes ``measuring'' the coupling constants at the Born--term poles, 
scanning the region around the origin of the Mandelstam plane in a search for 
the positions of the ``on--mass--shell avatars'' of the Adler zeros, and 
finally trying and getting at the values of the $\Sigma$ terms.

Of course, no one can hope (even with these techniques) to extract values for 
these parametres carrying relative errors better than those of the 
parametrizations used (which in turn are no better than the data they fit). 
For the moment, the only complete and adequate parametrisation (in the sense 
that it correctly describes the low--energy features of the data, including 
what one knows and/or expects to happen in the $\bar{K}N$ 
unphysical region \cite{VPI72}) 
is the old K--matrix analysis by Kim \cite{Kim67}.

\section*{5. Final remarks.}

The above is just an outline of work done within the EuroDA$\Phi$NE Network 
by the small fraction of it working on kaon--nucleon and kaon--nuclear 
physics. Of course it can only be consided as exploratory in absence of new 
data: but it has already helped in pointing out serious weaknesses in a field 
often thought and spoken of as well known and settled. It has since long been 
the learned opinion of the present authors that this is quite far from the 
truth \cite{KAON}; but we can now show that improvements are at hand, 
and some of them could already be implemented in a rather easy way, 
although with scarce expectation for success, in 
absence of new and better data.

\bibliographystyle{unsrt}

\end{document}